\documentclass{pasj00}

\begin{document}

\SetRunningHead{Shimasaku et al.}{Galaxies at $z \sim 6$}
\Received{2005/12/19}
\Accepted{2005/04/15}

\title{Number Density of Bright Lyman-Break 
Galaxies at $z\sim 6$ in the Subaru Deep Field
\altaffilmark{1}
}

\author{%
  Kazuhiro \textsc{Shimasaku} \altaffilmark{2,3},
  Masami   \textsc{Ouchi}     \altaffilmark{4,5},
  Hisanori \textsc{Furusawa}  \altaffilmark{6}
  Makiko   \textsc{Yoshida}   \altaffilmark{2}, \\
  Nobunari \textsc{Kashikawa} \altaffilmark{7}, 
  Sadanori \textsc{Okamura}   \altaffilmark{2,3}, 
}

\altaffiltext{1}
{Based on data collected at the Subaru Telescope, which is 
operated \\ by the National Astronomical Observatory of Japan.}

\altaffiltext{2}{Department of Astronomy, School of Science,
        The University of Tokyo, Tokyo 113-0033
        \\ Email (KS): shimasaku@astron.s.u-tokyo.ac.jp}
\altaffiltext{3}{Research Center for the Early Universe, 
        School of Science,
        The University of Tokyo, Tokyo 113-0033}
\altaffiltext{4}{Space Telescope Science Institute, 
        3700 San Martin Drive,
        Baltimore, MD 21218, USA}
\altaffiltext{5}{Hubble Fellow}
\altaffiltext{6}{Subaru Telescope, 
        National Astronomical Observatory of Japan, 
        650 N. A'ohoku Place, Hilo, 
        \\ HI 96720, USA}
\altaffiltext{7}{National Astronomical Observatory, 
        Mitaka, Tokyo 181-8588}
%
%

\maketitle

\begin{abstract}
We report on the bright Lyman-break galaxies (LBGs) 
selected in a 767 arcmin$^2$ area of the Subaru Deep Field.
The selection is made 
in the $i-z_{\rm R}$ versus $z_{\rm B}-z_{\rm R}$ plane, 
where $z_{\rm B}$ and $z_{\rm R}$ are new bandpasses with a central 
wavelength of 8842\AA\ and 9841\AA, respectively.
This set of bandpasses enables us to separate well $z \sim 6$ LBGs 
from foreground galaxies and Galactic cool stars.
We detect 12 LBG candidates down to $z_{\rm R}=25.4$, 
and calculate the normalization of the rest-frame 
far-ultraviolet (FUV: $\simeq 1400$\AA) luminosity function 
at $M_{\rm FUV} = -21.6$ to be 
$\phi(-21.6) = (2.6 \pm 0.7) \times 10^{-5}$ mag$^{-1}$ Mpc$^{-3}$.
This must be the most reliable measurement ever obtained 
of the number density of bright $z \sim 6$ LBGs, 
because it is more robust against both contamination and cosmic 
variance than previous values.
The FUV luminosity density contributed from 
LBGs brighter than $M_{\rm FUV} = -21.3$ is 
$(2.8 \pm 0.8) \times 10^{24}$ ergs s$^{-1}$ Hz$^{-1}$ Mpc$^{-3}$, 
which is equivalent to a star formation rate density 
of $(3.5 \pm 1.0) \times 10^{-4} M_\odot$ yr$^{-1}$ Mpc$^{-3}$.
Combining our measurement with those at $z<6$ in the literature, 
we find that the FUV luminosity density of bright galaxies 
increases by an order of magnitude from $z\sim 6$ to $\sim 3$ 
and then drops by $10^3$ from $z \sim 3$ to the present epoch, 
while the evolution of the total luminosity density is much milder.
The evolutionary behavior of bright LBGs resembles that 
of luminous dusty star-forming galaxies and bright QSOs.
The redshift of $z \sim 3$ appears to be a remarkable era
in the cosmic history when massive galaxies  
were being intensively formed.
\end{abstract}

%
%

\section{Introduction}

Over the last decade, 
selection of star-forming galaxies at high redshift 
using continuum breaks seen in their far-ultraviolet (FUV) 
spectra has been applied successfully 
to deep multi-color images taken with large telescopes.
Thousands of galaxies beyond $z \sim 2$ have been detected 
by this method in two-color planes; 
galaxies selected in this way are called Lyman-break galaxies (LBGs).
For instance, LBGs at $z \sim 4$ can be isolated from foreground 
objects (lower-redshift galaxies and Galactic stars)  
in the $B-R$ vs $R-i$ plane, 
since they have extremely red $B-R$ colors and relatively blue 
$R-i$ colors.
Various properties of LBGs up to $z \sim 5$ 
have been discussed in detail 
based on photometric data and follow-up spectroscopy  
(e.g., Steidel et al. 1999; Papovich et al. 2001; 
Shapley et al. 2001, 2003; 
Giavalisco \& Dickinson 2001; Porciani \& Giavalisco 2002; 
Foucaud et al. 2003; 
Iwata et al. 2003; Ando et al. 2004; Ouchi et al. 2004a,b).

Very recently, several groups have made surveys of 
a similar galaxy population at $z \sim 6$ 
using ultra-deep imaging data taken 
with Hubble Space Telescope (HST) and 8m-class telescopes.
Most of them search for galaxies red in $i-z$. 
Such {\lq}$i$-dropout{\rq} galaxies are 
likely star-forming galaxies at $z \sim 6$, since 
the continuum break of $z \sim 6$ galaxies falls into the $i$ band.
Systematic studies of galaxies at $z \sim 6$ and beyond are 
crucial not only for the understanding of the formation and 
early evolution of galaxies during and just after the reionization 
epoch of the universe, but also for the identification of 
objects which contributed to the reionization.

We begin with a brief summary of the previous studies 
on $z \sim 6$ galaxies.
Fontana et al. (2003) made a multicolor Very Large Telescope (VLT) 
and HST imaging survey aiming at detecting $4.5<z<6$ galaxies.
In their VLT data for an area of about 30 arcmin$^2$, 
13 candidates with $Z_{\rm AB} \le 25$ mag were found based on 
photometric redshifts, among which four were at $5 < z \le 6$.
Dickinson et al. (2004) detected 
in early Advanced Camera for Surveys (ACS) GOODS data 
of 316 arcmin$^2$ five $z \sim 6$ candidates with $i-z > 1.3$ 
down to $z \simeq 26$ mag.
Stanway et al. (2003, 2004) detected eight and six 
$i$-dropout ($i-z > 1.5$) galaxies with $z \le 25.6$ 
in public HST ACS data of the GOODS-South (146 arcmin$^2$ area) 
and the GOODS-North (200 arcmin$^2$), respectively.
Bouwens et al. (2003) detected 23 $z \sim 6$ galaxy candidates 
in two ACS Guaranteed Time Observation fields (46 arcmin$^2$ in total) 
down to $z \sim 27.3$ mag.
A deeper survey was made subsequently by Bouwens et al. (2004a) 
for two Hubble Ultra Deep parallel fields; 30 $i$-dropouts 
($i-z > 1.4$) were found over 21 arcmin$^2$ down to $z=28.1$.
Bouwens et al. (2004b) discussed the size evolution 
of galaxies up to $z \sim 6$ in the Hubble Ultra Deep Field (HUDF) 
and the Hubble Ultra Deep parallel fields.
Yan, Windhorst, \& Cohen (2003) found 30 $z \sim 6$ candidates 
brighter than $z \simeq 28$ mag in an ACS/WFC parallel field 
(10 arcmin$^2$).
Yan \& Windhorst (2004) found 108 candidates 
with $z \le 30.0$ mag in the HUDF (10.34 arcmin$^2$).
Bunker et al. (2004) also made a search in the HUDF, 
finding 54 candidates at $z < 28.5$ mag with $i-z > 1.3$.
Many of the studies have reported 
a decrease in the FUV luminosity density of bright 
galaxies by a factor of $\approx 3-8$ from $z \sim 3$ to $z \sim 6$ 
(Dickinson et al. 2004; Giavalisco et al. 2004; 
Stanway et al. 2003, 2004; 
Bouwens et al. 2004a; Bunker et al. 2004), 
although Fontana et al. (2003) and Bouwens et al. (2003) obtained 
a roughly constant luminosity density over this redshift range. 
Measuring the number density of bright LBGs accurately 
is of great importance in discussing and modeling 
the formation and evolution of massive galaxies.

There are, however, two main uncertainties involved
in the previous studies of bright $z \sim 6$ LBGs.
First, selection of LBGs by a single $i-z$ color is likely 
to suffer from a significant amount of contamination, 
especially at bright magnitudes, 
since red galaxies at $z \sim 1$ -- $2$ and Galactic cool stars 
can be as red in $i-z$ as $z \sim 6$ LBGs.
Removal of unresolved objects in a high-resolution HST image 
is an efficient way to reduce the contamination by cool stars. 
However, high-resolution imaging is not helpful 
to separate red $z \sim 1$ -- $2$ galaxies from LBGs. 
Near infrared (NIR) photometry is useful especially for 
discriminating LBGs from red galaxies. 
In the previous studies, however, NIR photometry was either 
limited to a very small sky area, or not deep enough to securely 
separate red galaxies from LBGs.
Second, the areas surveyed in the previous $z \sim 6$ 
LBG papers are a few tens to a few hundreds arcminutes, 
the widest being 346 arcmin$^2$. 
This may not be sufficiently large when one recalls  
an extremely low surface density of bright LBGs 
(an order of $\sim 10^{-2}$ arcmin$^{-2}$) and 
large cosmic variance in their volume density expected 
from observations of lower-redshift LBGs 
and biased galaxy formation models.
Thus, wide-field and deep two-color imaging is essential 
for constructing a large and clean sample of $z \sim 6$ LBGs.

In this paper, we report on a survey of $z \sim 6$ LBGs which is
aimed to overcome the two main uncertainties in the previous
surveys. We apply for the first time a two-color selection
to imaging data as wide as 767 arcmin$^2$. We obtain a
reliable value of the number density of bright $z \sim 6$ LBGs.
By combining our value with those in the literature, we
examine the evolution of the number density of bright LBGs
over $3 < z < 6$ and discuss the formation of massive galaxies.

The plan of this paper is as follows.
Section 2 explains the strategy of our survey.
Section 3 describes the photometric data used in our analysis.
Selection of $z \sim 6$ LBGs is made in \S 4.
Section 5 is devoted to results and discussion, 
and a conclusion is given in \S 6.
AB magnitudes are used in this paper.
We assume a flat universe with $\Omega_0=0.3$, $\lambda_0=0.7$, 
and $H_0 = 70$ km s$^{-1}$ Mpc$^{-1}$.

%
%

\section{Strategy of Our Survey}

Our survey covered an area of 767 arcmin$^2$ of the Subaru
Deep Field (SDF: Maihara et al. 2001).
This wide survey area will considerably smooth out large-scale
cosmic variance which will be probably at present at $z \sim 6$.
Several observations have reported large cosmic variance
for $z > 5$ galaxies. 
Bremer et al. (2004) found that the surface
density of $z \sim 5$ galaxies varies considerably over areas
comparable to a single ACS pointing. 
Shimasaku et al. (2003, 2004) reported large inhomogeneity of 
the sky distribution of Lyman $\alpha$ emitting galaxies at $z \sim 5$ 
on scales of 50 Mpc.
At $z \sim 6$, a transverse comoving length of 50 Mpc 
corresponds to $20'$. 
Hu et al. (2004) and Ouchi et al. (2005) also found
a highly inhomogeneous spatial distribution of $\simeq 5.7$ Lyman 
$\alpha$ emitting galaxies on tens of Mpc scales. 
See Somerville et al. (2004) for a theoretical prediction for 
cosmic variance of high-redshift galaxies.

\begin{figure}
  \hspace{-45pt}
  \FigureFile(120mm,120mm){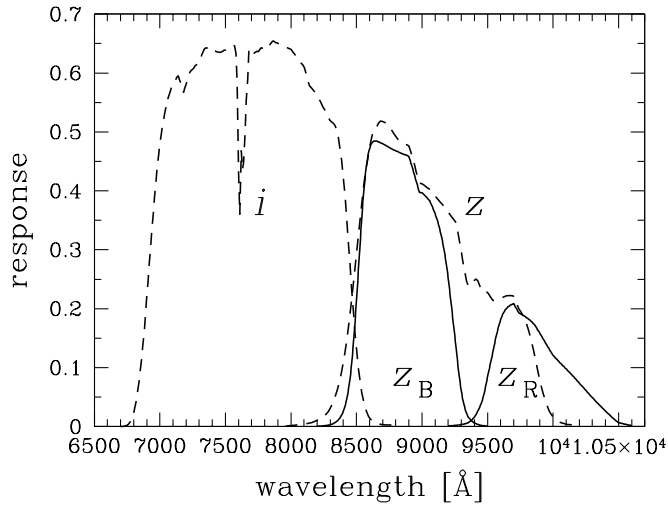}
  \vspace{-20pt}
  \caption{
         Response functions of the $z_{\rm B}$ and $z_{\rm R}$ 
         bandpasses (solid lines).
         The two dashed lines indicate the response of 
         the $i$ and $z$ bands, respectively, 
         of the Subaru/Suprime-Cam.
    \label{fig:zbzr}}
\end{figure}

The bandpasses we use for two-color selection are $i$, $z_{\rm B}$,
and $z_{\rm R}$. 
The $z_{\rm B}$ and $z_{\rm R}$ filters, which were designed by us 
for this purpose, divide the SDSS $z$ band into two at 9500 \AA. 
Figure \ref{fig:zbzr} plots the effective response curves of 
the $z_{\rm B}$ and $z_{\rm R}$ bandpasses which include the filter 
transmission, CCD sensitivity, throughput of the optics of the 
Subaru Telescope, and atmospheric extinction at an airmass of 1.2. 
The central wavelength and FWHM of the $z_{\rm B}$ and
$z_{\rm R}$ bandpasses are (8842\AA, 689\AA) and (9841\AA, 537\AA),
respectively. 
These two bandpasses enable us to measure the slope of continuum 
emissions redward of Lyman $\alpha$, which is crucial
for discriminating $z \sim 6$ objects from $z \sim 1$ -- $2$  
red galaxies and cool stars both of which have similar 
$i - z_{\rm R}$ but redder $z_{\rm B} - z_{\rm R}$ colors (see \S 4).

We have another three advantages of using $z_{\rm B}$ and $z_{\rm R}$. 
First is that we can measure reliably rest-frame UV continuum 
flux from $z_{\rm R}$ photometry.
The continuum flux of $i-z$ selected LBG candidates 
is usually measured from their $z$-band magnitude.
However, a recent study based on spectroscopic observation 
of an $i$-dropout galaxy at $z=6.33$ suggests that a large 
fraction of $i$-dropout galaxies may have 
a strong Ly$\alpha$ emission (Nagao et al. 2004). 
If this is true, measurements of the continuum flux of $z>6$ 
galaxies from $z$-band photometry would be biased high 
due to contamination from Ly$\alpha$ emission.
A galaxy's Lyman $\alpha$ line does not enter the $z_{\rm R}$ bandpass 
as long as the galaxy is at $z<6.7$.

Second, we can construct an LBG sample of a well-defined 
and relatively narrow 
redshift range thanks to $z_{\rm B}-z_{\rm R}$ color.
As will be shown in \S 4, the $z_{\rm B}-z_{\rm R}$ value of 
star-forming galaxies increases rapidly with redshift at $z>6$.
Thus, selection of LBG candidates with $z_{\rm B}-z_{\rm R}$ color less 
than an appropriate value ($z_{\rm B}-z_{\rm R} = 0.7$ in this study) 
provides an LBG sample of a redshift distribution 
with a sharp upper limit, 
although the precise correspondence between $z_{\rm B}-z_{\rm R}$ and 
redshift varies with the intrinsic spectral shape of LBGs.
This is important in calculating the comoving volume surveyed by 
our observation.
In contrast, the selection function of a single $i-z$ color 
criterion is wide, having a long high-redshift tail.

Third, our $z \sim 6$ LBG sample is free from contamination 
by spurious objects near the detection limit of the $z_{\rm R}$ data.
As will be seen in \S 4, one of our selection criteria 
for $z \sim 6$ LBGs is $z_{\rm B}-z_{\rm R} \le 0.7$. 
Since the $z_{\rm B}$ data are deeper than the $z_{\rm R}$ data 
by about 0.7 mag, this criterion ensures that objects satisfying 
this criterion are detected in the $z_{\rm B}$ image as well.
Fake objects in the $z_{\rm R}$ image which happen to exceed 
the detection flux are basically undetected 
in the $z_{\rm B}$ image, and thus will be eliminated by this criterion.
Selection by $i-z$ color alone could be contaminated by spurious 
objects near the detection limit.

%
%

\section{Data}

\subsection{$z_{\rm B}$ and $z_{\rm R}$ Imaging}

We took deep $z_{\rm B}$- and $z_{\rm R}$-band imaging data 
of a $30'\times 24'$ area in the Subaru Deep Field (SDF) 
[$13^{\rm h} 24^{\rm m} 38.^{\rm s}9$, $+27^\circ 29 ' 26''$(J2000)] 
with the prime focus camera (Suprime-Cam; Miyazaki et al. 2002)
mounted on the 8.2m Subaru telescope 
on 2003 July 2 and 2004 March 19 and 20.
The image scale of Suprime-Cam is $0.''202$ per pixel.
The exposure time was 85 min for $z_{\rm B}$ and 599 min 
for $z_{\rm R}$;
we set a much longer exposure time for $z_{\rm R}$, 
since the effective throughput of the $z_{\rm R}$ band 
is considerably lower than that of $z_{\rm B}$ 
due to the rapidly decreasing quantum efficiency of the CCDs toward 
longer wavelengths beyond 9500 \AA.
The individual CCD data were reduced and
combined by IRAF and a data reduction software package 
for Suprime-Cam developed by Yagi et al. (2002) and Ouchi (2003).
The combined images for individual bands were aligned and smoothed 
with Gaussian kernels so that the PSF sizes of the final images 
match those of the public images described in the next subsection.
The final images after removal of low $S/N$ regions near the 
edges of the field of view cover a contiguous
767 arcmin$^2$ area with a PSF FWHM of $0.''98$.
Photometric calibration was made using 
the spectrophotometric standard star 
GD153 (Bohlin, Colina, \& Finley 1995) 
taken on 2003 July 2 for both bandpasses.
A summary of the data is given in Table \ref{tab:obsdata}.

\begin{table}
  \caption{Summary of the observational data.
    \label{tab:obsdata}
  }
  \begin{center}
    \begin{tabular}{cccl}
\hline
\hline
Band & Exposure [min] & $m_{\rm lim}^{\ast}$ & Obs. date \\
\hline
$z_{\rm B}$ &  85 & 26.2 & 2 July 2003 \\
$z_{\rm R}$ & 599 & 25.4 & 2 July 2003 \\
            &     &      & 19, 20 March 2004 \\
$B$   & 595 & 28.5 & public \\
$V$   & 340 & 27.7 & public \\
$R$   & 600 & 27.8 & public \\
$i$   & 801 & 27.4 & public \\
$z$   & 504 & 26.6 & public \\
NB816 & 600 & 26.6 & public \\
NB921 & 899 & 26.5 & public \\
\hline
    \end{tabular}
  \end{center}

\noindent
$\ast$ $3\sigma$ limiting magnitudes on a $2''$ diameter aperture.

\end{table}

\subsection{Public Imaging Data}

The SDF has deep, public Suprime-Cam data of seven bandpasses, 
$B$, $V$, $R$, $i$, $z$, NB816, and NB921, 
obtained in the Subaru Deep Field Project (Kashikawa et al. 2004).
This project is a large program of Subaru Observatory to carry out 
a deep galaxy survey in a blank field in optical and NIR 
wavelengths to study distant galaxies.
$B$, $V$, $R$, $i$, and $z$ are standard Johnson and SDSS 
broad-bands, and NB816 and NB921 are narrow-bands 
whose central wavelength and FWHM are 
(8150 \AA, 120 \AA) and (9196 \AA, 132 \AA), respectively.
The exposure times and limiting magnitudes of the public images 
are given in Table \ref{tab:obsdata}.
We combine the $i$-band data with our $z_{\rm B}$ and $z_{\rm R}$ data 
to select LBGs at $z\sim 6$ 
in the $i-z_{\rm R}$ vs $z_{\rm B}-z_{\rm R}$ plane. 
We also use the $B,V,R$ data to further remove 
foreground objects from the LBG sample.

\subsection{Object Detection and Photometry}

\begin{figure}
  \hspace{-20pt}
  \FigureFile(110mm,110mm){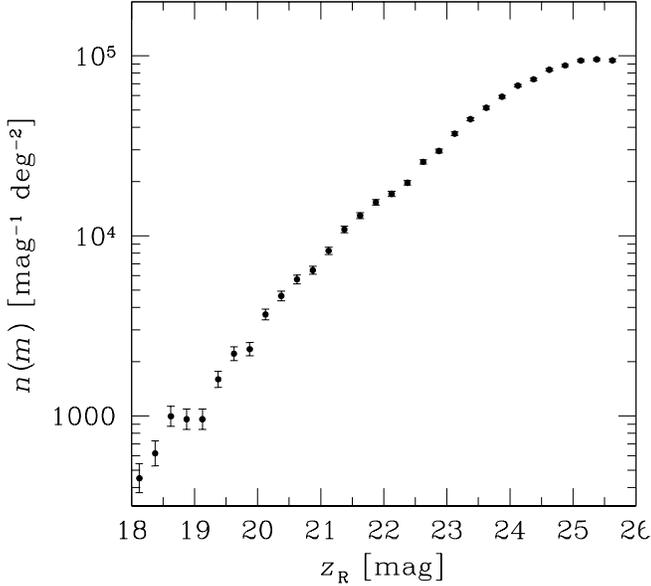}
  \vspace{-20pt}
  \caption{
         Number counts of galaxies in the $z_{\rm R}$band.
         Detection incompleteness has not been corrected.
         The error bars imply Poisson errors.
    \label{fig:nm}}
\end{figure}

Object detection and photometry were 
made using SExtractor version 2.1.6 (Bertin and Arnouts 1996) 
on all nine images ($z_{\rm B}$, $z_{\rm R}$, and seven public images).
The $z_{\rm R}$-band image was chosen to detect objects. 
If more than 5 pixels whose counts were above the 2 $\sigma_{\rm sky}$ 
were connected, they were regarded as an object.
In total, 45,405 objects were detected down to $z_{\rm R}=25.4$ mag 
($3\sigma$ limiting magnitude).
Figure \ref{fig:nm} plots the number counts of objects 
detected in the $z_{\rm R}$ band, where stellar objects 
with the PSF shape brighter than $z_{\rm R}=23$ have been removed.
For each object detected in the $z_{\rm R}$ image, 
a magnitude on a $2''$-diameter aperture was measured for 
each passband to derive the colors of the object;  
we adopted for each object the same aperture among all passbands 
to ensure that the colors were measured correctly.
We adopted MAG$\_$AUTO for the total $z_{\rm R}$ magnitude.

%
%

\section{Selection of $z\sim 6$ Lyman-break Galaxies}

\begin{figure}
  \hspace{-20pt}
  \FigureFile(100mm,100mm){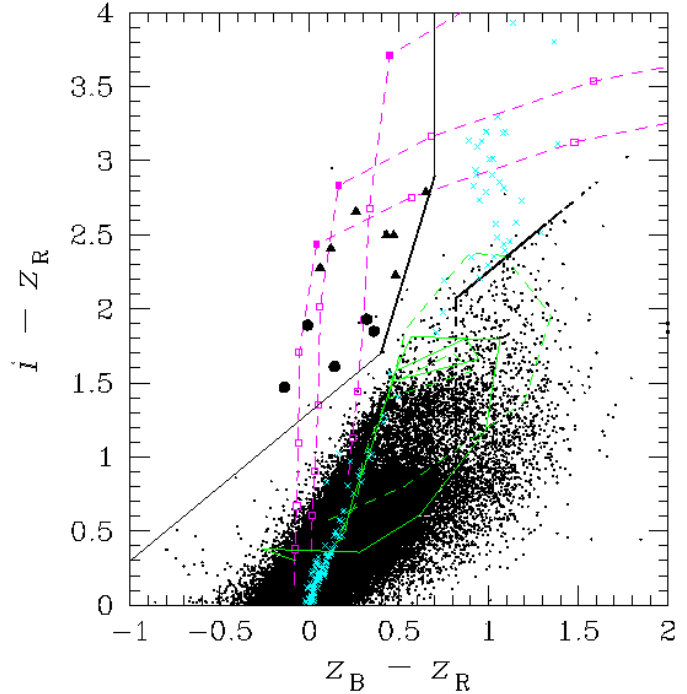}
  \caption{
         Distribution of objects 
         in the $i-z_{\rm R}$ versus $z_{\rm B}-z_{\rm R}$ plane.
         The dots indicate objects detected in our data. 
         If the $i$ or $z_{\rm B}$ magnitude of an object is fainter 
         than the $2\sigma$ magnitude of the corresponding bandpass, 
         it is replaced with the $2\sigma$ magnitude.
         The dots along the line 
         $i-z_{\rm R} = z_{\rm B}-z_{\rm R}+1.25$ 
         at $z_{\rm B}-z_{\rm R}>0.8$ are objects which are fainter 
         than the $2\sigma$ magnitude in both $i$ and $z_{\rm B}$.
         The 12 LBG candidates are shown by filled circles 
         ($i$ magnitudes brighter than the $2\sigma$ mag) and 
         filled triangles (fainter than the $2\sigma$ mag).
         The magenta dashed lines represent the predicted 
         tracks of young star-forming galaxies with three 
         different dust extinction values: 
         from lower left to upper right, $E(B-V)=0, 0.15, 0.5$. 
         On the tracks, redshifts are marked by small squares 
         with 0.2 interval; the filled squares correspond to $z=6$.
         The green lines imply the tracks of elliptical galaxies 
         redshifted from $z=0$ to 3 without evolution; 
         the spectra are taken from 
         Fukugita et al. (1995; solid line) and 
         Coleman et al. (1980; dashed line).
         The cyan crosses represent Galactic stars given in 
         Gunn \& Stryker (1983) and Knapp et al. (2004).
         The adopted boundary for LBG selection is shown by 
         the thick solid line.
    \label{fig:twocolor}}
\end{figure}

Figure \ref{fig:twocolor} shows the selection of 
LBGs at $z \sim 6$ in the $i-z_{\rm R}$ vs $z_{\rm B}-z_{\rm R}$ plane.
The magenta curves in this figure indicate the tracks of young, 
star-forming galaxies at $z \ge 5$ 
with three dust extinction values, $E(B-V)=0, 0.15, 0.50$; 
the filled squares on the curves correspond to $z=6$.
The value of $E(B-V)=0.15$ is the median extinction of 
$z\sim 4$ LBGs detected in the SDF (Ouchi et al. 2004a).
For the intrinsic spectrum of the young galaxies, we adopt 
the spectral synthesis model of Kodama \& Arimoto (1997) 
and set an age of 0.1 Gyr, Salpeter IFM, and a star-formation 
timescale of 5 Gyr.
We use Calzetti et al.'s (2000) formula for dust extinction 
and Madau's (1995) prescription for the absorption by the IGM.
It is found that the $i-z_{\rm R}$ colors of the model galaxies 
increase monotonically with redshift up to $z=6$ 
while their $z_{\rm B}-z_{\rm R}$ colors stay nearly constant.
At $z>6$, the model galaxies become rapidly redder 
in $z_{\rm B}-z_{\rm R}$.

The crosses represent 175 Galactic stars of a wide variety of types 
given in the spectrophotometric atlas of Gunn \& Stryker (1983) 
and 46 Galactic cool stars taken from Knapp et al. (2004) 
for which spectra are available down to 7000 \AA.
The green solid and green dashed curves represent the tracks 
of elliptical galaxies from $z=0$ to $z=3$ (without evolution) 
given in Fukugita, Shimasaku, \& Ichikawa (1995)
and Coleman, Wu, \& Weedman (1980), respectively; 
these two curves are plotted 
to show the position and its typical uncertainty 
in the $i-z_{\rm R}$ vs $z_{\rm B}-z_{\rm R}$ plane 
of the reddest interlopers.
Coleman et al.'s spectrum does not show a UV upturn and thus 
has redder colors than Fukugita et al.'s. 

Figure \ref{fig:twocolor} demonstrates that LBGs 
at $z \simeq 5.6$ -- $6.2$ are 
separated well from foreground objects (intermediate-redshift 
red galaxies and Galactic stars) 
in the $i-z_{\rm R}$ vs $z_{\rm B}-z_{\rm R}$ plane.
A simple selection of LBGs using a single color, e.g., $i-z_{\rm R}$, 
may suffer from contamination, 
since red galaxies at $z\sim 1$ -- $2$ and cool stars have 
similar $i-z_{\rm R}$ colors to $z \sim 6$ LBGs; 
$z_{\rm B}-z_{\rm R}$ color plays an essential role 
in discriminating such interlopers from LBGs.

Previous two-color selections of $z \lesssim 5$ LBGs have 
usually adopted for the vertical axis of the two-color plane 
(i.e., the axis sensitive to the Lyman break) 
the bluest bandpass minus the middle-wavelength bandpass; 
for example, for $B,R,i$-selected LBGs at $z \sim 4$, 
$B-R$, not $B-i$, is adopted as the vertical axis.
In this study, however, we have adopted $i-z_{\rm R}$, 
not $i-z_{\rm B}$, 
for the vertical axis from the technical reason that 
since we are based on a $z_{\rm R}$-detected catalog, some objects 
are undetected in $z_{\rm B}$, for which $i-z_{\rm B}$ colors cannot be 
expressed properly.
Of course, adopting $i-z_{\rm B}$ instead of $i-z_{\rm R}$ does not 
change our result as long as we use the same selection criteria.

From Figure \ref{fig:twocolor}, 
we define the selection criteria for LBGs at $z \sim 6$ as:
\begin{eqnarray}
i-z_{\rm R} \ge z_{\rm B} - z_{\rm R} + 1.3,    \nonumber \\ 
z_{\rm B} - z_{\rm R} \le 0.7,                  \nonumber \\
i-z_{\rm R} \ge 4(z_{\rm B} - z_{\rm R}) + 0.1, \nonumber \\
B,V,R \ge 2\sigma \hspace{5pt} {\rm mag}.       \nonumber 
\end{eqnarray}
The selection boundary in the $i-z_{\rm R}$ vs $z_{\rm B}-z_{\rm R}$ 
plane defined by these criteria is shown by the black solid lines 
in the figure.
These criteria select a model galaxy with $E(B-V)=0.15$ at 
$z=5.6$ -- $6.2$.
Varying $E(B-V)$ over 0 and 0.5 does not significantly 
change the redshift range selected.
To further reduce contamination from foreground galaxies, 
we impose the fourth criterion that an LBG candidate must 
be un-detected, i.e., less than the $2\sigma$ limiting flux, 
in any of the $B$, $V$, and $R$ images.

The dots indicate the objects detected in our data. 
The black circles and triangles indicate LBG candidates 
in our data which satisfy all the selection criteria;
triangles correspond to objects not detected in $i$, 
i.e., fainter than the $2\sigma$ magnitude.
All the observed magnitudes and colors have been corrected 
for Galactic absorption using Schlegel, Finkbeiner, \& Davis (1998).

We select 12 LBG candidates in total down to $z_{\rm R}=25.4$, 
which are summarized in Table \ref{tab:lbgs}.
We do not plot the error bars in Figure \ref{fig:twocolor} 
to avoid crowding. 
The typical $1\sigma$ errors in $z_{\rm B}-z_{\rm R}$ and 
$i-z_{\rm R}$ 
for the 12 candidates are $0.3$ mag and $0.4$ mag, respectively
(The $i-z_{\rm R}$ error is estimated for objects 
with $i$ magnitudes brighter than the $2\sigma$ mag).
The total $z_{\rm R}$ magnitudes (MAG\_AUTO) of all 12 candidates 
are fainter than 24.8.
However, all the candidates are visible in both $z_{\rm B}$, $z$, 
and NB921 images above the $3\sigma$ level, meaning that 
they are not spurious objects.

The FWHMs of the 12 candidates in the $z_{\rm R}$ image 
span over $0.''86$ - $1.''78$, with a median of $1.''2$ 
and an interquartile range of $1.''14$ -- $1.''53$.
Similar FWHM values are obtained for the $z$ image 
which is considerably deeper than $z_{\rm R}$.
For most of them, the deviation from the PSF is not significant, 
when we consider that SExtractor tends to return FWHM values 
slightly larger than the PSF size for very faint point sources.
Visual inspection of the 12 candidates in the $z$ image 
finds that three objects seem to be resolved, 
with a major axis of about $2''$.
The relatively large seeing size of our data prevents us from 
measuring the intrinsic sizes of candidates, 
but some brightest LBGs at $z \sim 6$ could 
have a large star forming region of several kpc.

\begin{figure}
  \FigureFile(90mm,90mm){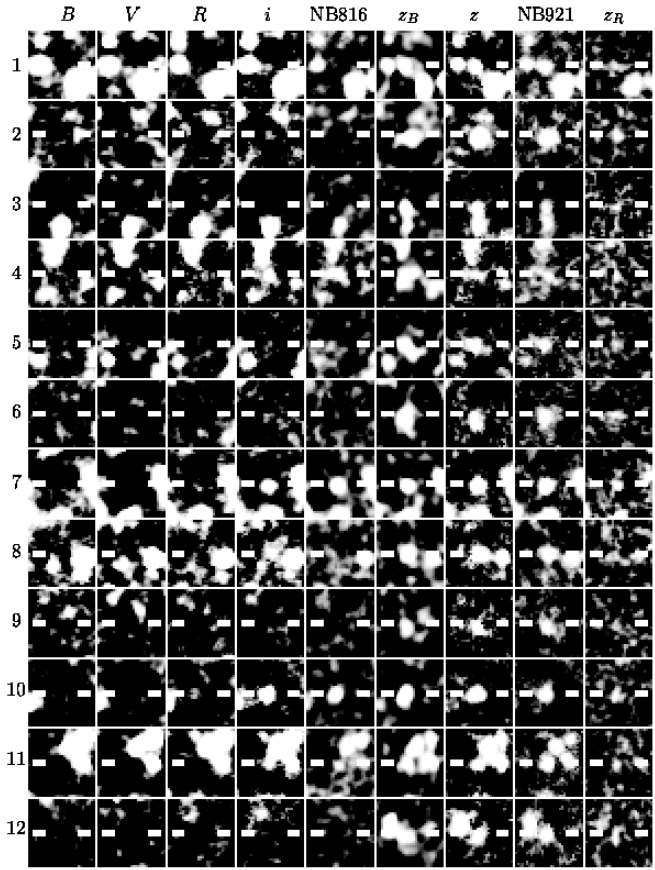}
  \vspace{0pt}
  \caption{
         Snapshots of 12 LBG candidates. 
         From left to right, 
         $B,V,R,i$,NB816,$z_{\rm B},z$,NB921,$z_{\rm R}$ 
         images are placed.
         The image size is $6''$ a side.
         North is up and east is to the left.
    \label{fig:snapshot}}
\end{figure}

\begin{figure}
  \hspace{-20pt}
  \FigureFile(110mm,110mm){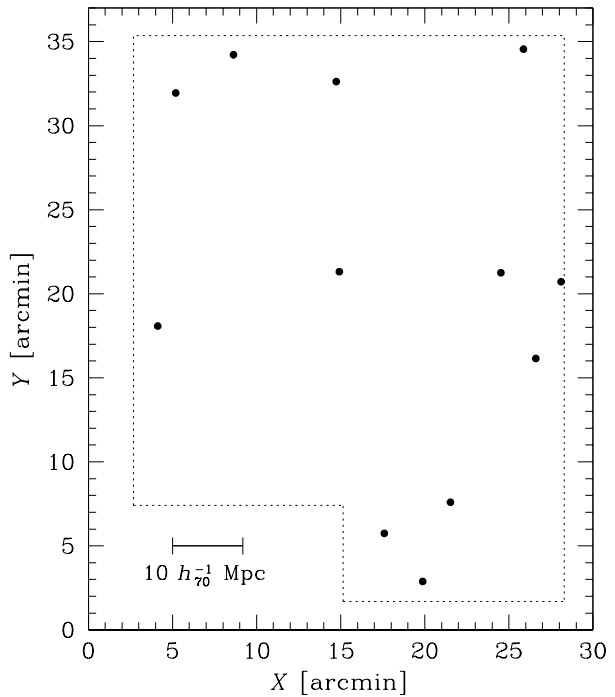}
  \caption{
         Sky distribution of 12 LBG candidates.
         LBG selection is made in the region outlined 
         by the dotted line.
    \label{fig:xy}}
\end{figure}

Snapshots of the 12 candidates are shown in Figure \ref{fig:snapshot}.
In this figure, the ID number increases from top to bottom, 
and the bandpasses are
$B, V, R, i, {\rm NB816}, z_{\rm B}, z, {\rm NB921}, z_{\rm R}$ 
from left to right.
All the objects are found to be 
detected clearly in both $z_{\rm B}$, $z$, and NB921, 
while some are not visible in $i$ or NB816, 
which are expected to be at higher redshifts.
The sky distribution of the 12 candidates is shown 
in Figure \ref{fig:xy}. 
The dotted lines outline the region 
in which the LBG selection is made. 
The area in the lower-left corner, which corresponds in size to 
the field of view of a single CCD, is not used in this study, 
since the quantum efficiency of the CCD in charge of this area 
is about two thirds those of the other nine CCDs 
and thus the data of this area are considerably noisy.

We estimate a possible contamination rate for simple 
$i$-dropout selection.
If we omit $z_{\rm B} - z_{\rm R} \le 0.7$ and 
$i-z_{\rm R} \ge 4(z_{\rm B} - z_{\rm R}) + 0.1$ 
criteria and thus select objects based on a simple $i$-dropout 
criterion ($i-z_{\rm R} \ge z_{\rm B}-z_{\rm R}+1.3$, 
similar to the criteria adopted in the previous studies) 
and non-detection in $B,V,R$, then we have ten more candidates.
These newly selected ten objects might be LBGs at $z > 6.2$ 
or very dusty LBGs at $z < 6$, 
but they are more likely to be foreground objects.
If all of them are really foreground objects, 
the contamination fraction will be $\simeq 50\%$, 
implying that bright single-color selected $i$-dropout samples 
for $z \sim 6$ LBGs are highly contaminated.
An accurate evaluation of contamination needs spectroscopic 
observation of these objects. 
It should be, however, emphasized that the two criteria we omit 
in this simulation, 
which set an upper limit to the $z_{\rm B}-z_{\rm R}$ 
color of LBG candidates,  
not only reduce a possible contamination from foreground objects 
but also provide a sharp upper limit to the redshift range of 
our LBG sample.

Dickinson et al. (2004) found five $z \sim 6$ candidates 
in HST ACS data of the two GOODS fields 
based on $i-z$ color. 
Among them, three are brighter than $z = 25.4$, 
giving a surface density of 0.009 arcmin$^{-2}$.
Stanway et al. (2003, 2004) made a similar 
$i$-drop survey in the same fields.
There are in total nine LBG candidates with $z \le 25.4$ 
in their LBG samples; 
the surface density is 0.026 arcmin$^{-2}$.
There is one object with $z \le 25.4$ in Bouwens et al.'s (2003) 
data of the ACS Guaranteed Time Observation fields (46 arcmin$^2$), 
giving a surface density of 0.02 arcmin$^{-2}$.
These values differ by up to a factor of two from ours, 
12 candidates / 767 arcmin$^2$ = 0.016 arcmin$^{-2}$. 
In general, however, surface density measurements depend 
on selection criteria, redshift range surveyed, 
and detection completeness, 
which are different among the authors.
Thus, we do not make a detailed comparison of surface density. 

%
%

\section{Results and Discussion}

\subsection{Number Density of Bright LBGs}

Since the 12 LBG candidates are distributed in a narrow range 
of $z_{\rm R}$ magnitude, $z_{\rm R}=24.8$ -- $25.4$, 
we do not fit the Schechter 
function to the data to derive the best-fit values of the three 
Schechter parameters ($\alpha$, $M^\star$, and $\phi^\star$).
Instead, we measure the normalization of the luminosity function 
(num mag$^{-1}$ Mpc$^{-3}$) 
at the average $z_{\rm R}$ magnitude of our sample ($z_{\rm R}=25.1$).
This average magnitude corresponds to an rest-frame 
far-ultraviolet (FUV: around 1400\AA) absolute magnitude 
of $M_{\rm FUV} = -21.6$ mag at $z=5.9$, 
the central redshift of our survey.
The absolute magnitude is calculated as 
$M_{\rm FUV} = m - 5\log(d_L(z)) + 2.5\log(1+z)$, 
where $d_L(z)$ is the luminosity distance.
The rest-frame wavelength corresponding to the central 
wavelength of the $z_{\rm R}$ band changes from 1491\AA\ (at $z=5.6$) 
to 1367\AA\ ($z=6.2$). 
We assume that LBGs have flat spectra 
($f_\nu=$ const) over this wavelength range.

The value of the luminosity function at $M_{\rm FUV} = -21.6$ 
is calculated as:
\begin{equation}
\phi(-21.6) \ [{\rm mag}^{-1} {\rm Mpc}^{-3}]
= {1\over{0.6}} \times \sum_{i=1}^{12} {1\over{V_{\rm eff}(m_i)}}
\end{equation}
where $m_i$ is the $z_{\rm R}$ magnitude of the $i$-th candidate, 
$V_{\rm eff}(m_i)$ is the effective survey volume (comoving) for 
the $i$-th candidate, and 0.6 is the size of the magnitude bin. 
The effective volume for a given apparent magnitude $m$ is computed 
by $V_{\rm eff}(m) = \int_z p(m,z) {dV\over{dz}} dz$, 
where $p(m,z)$ is the probability that an LBG of 
magnitude $m$ and redshift $z$ is detected 
in our $z_{\rm R}$ image and passes the selection criteria defined 
above when photometric errors are taken into account 
(Steidel et al. 1999), 
and ${dV\over{dz}}$ is the differential comoving volume 
for a solid angle of 767 arcmin$^2$.

We estimate $p(m,z)$ assuming that the color distribution of 
$z \sim 6$ LBGs is similar to that of bright $z \sim 4$ LBGs 
detected in the SDF by Ouchi et al. (2004a) 
and that color differences among LBGs are due to different 
$E(B-V)$ values.
Ouchi et al. (2004a) found that the $E(B-V)$ of $z \sim 4$ LBGs 
is distributed rather uniformly over $0 \le E(B-V) \le 0.3$ 
(They adopted a single, young age for their LBGs 
to estimate $E(B-V)$).
We generate a number of young LBGs uniformly 
over $0 \le E(B-V) \le 0.3$, 
assign them apparent $z_{\rm R}$ magnitudes, 
and distribute them randomly on the original $z_{\rm R}$ image.
The value of $p(m,z)$ is calculated as the ratio of the number 
of simulated LBGs (with given $z_{\rm R}$ mag and redshift $z$) 
which are detected and pass the selection criteria defined above, 
to the total number of simulated LBGs. 
In this simulation, we assume that the shape of LBGs is PSF-like 
(FWHM $= 1''.0$).
The systematic difference between input magnitudes
and magnitudes measured by SExtractor is found to be 
negligible in our simulation. 
The median FWHM value of simulated objects measured by SExtractor 
is also found to agree with that of the 12 LBG candidates.
The probability for $z_{\rm R}=25.1$, $p(25.1,z)$, 
is a function of redshift with a peak of $0.70$ at $z=5.9$ 
and a rather steep slope on both low-redshift and high-redshift 
sides; its FWHM is about 0.6.
Hence, the relation between apparent magnitude and absolute 
magnitude is sufficiently tight in our sample.

We obtain 
$\phi(-21.6) = (2.6 \pm 0.7) \times 10^{-5}$ mag$^{-1}$ Mpc$^{-3}$, 
where the error indicates the Poisson error ($=1/\sqrt{12}$).
Although we have not searched for the best-fit values 
for the Schechter parameters, the fact that no LBG candidate 
is found at $z_{\rm R}<24.8$ in our data appears 
to favor a faint $M_{\rm FUV}^\star$. 
About four LBGs with $z_{\rm R}<24.8$ are expected to be found 
if we fix $\alpha=-1.6$ and $M_{\rm FUV}^\star=-21.04$, 
which are the values for $z \sim 3$ LBGs 
($\phi^\star$ is adjusted so that $\phi(-21.6)$ takes 
the correct value).
Similarly, adopting $M_{\rm FUV}^\star=-20$ and $-22$ 
gives 0.7 and 9 candidates, respectively.
In the previous studies, Bouwens et al. (2004a) have 
obtained relatively faint $M_{\rm FUV}^\star = -20.26$ 
from their $i$-drop sample, 
while Bunker et al. (2004) found 
that the $M^\star$ of $z \sim 6$ LBGs is consistent with 
the value observed at $z \sim 3$.

Bouwens et al. (2004a) derived the luminosity function of 
$i$-drop galaxies at $z\sim 6$ from HST ACS data of the GOODS 
fields and the Hubble Ultra Deep parallel fields. 
The normalization of the bright part ($M_{\rm FUV}<-19.7$) 
of their luminosity function has been determined 
primarily from the wide-field ($>300$ arcmin$^2$) GOODS data.
The set of the best-fit Schechter parameters they obtained is 
$\alpha=-1.15$, $M_{\rm FUV}^\star=-20.26$, 
and $\phi^\star=0.00173$ Mpc$^{-3}$, which gives 
$\phi(-21.6) = 4.3 \times 10^{-5}$ Mpc$^{-3}$.
This $\phi(-21.6)$ value is about 70\% as high as the value 
obtained by us.
The major portion of this difference can be accounted for 
by the Poisson errors in both measurements, 
but cosmic variance probably contributes to it to some degree.
In any case, our number-density measurement must be more reliable, 
since it is based on wide-field two-color selection.
Our survey area is larger than 
the sum of the two GOODS fields by more than factor two, 
thus being less sensitive to cosmic variance.
Cosmic variance is probably higher for brighter LBGs, 
since observations of LBGs at $z \lesssim 5$ have revealed that 
brighter objects tend to be clustered more strongly 
(e.g., Giavalisco \& Dickinson 2001; Ouchi et al. 2004b).
We obtain a rough estimate of the cosmic variance following 
Somerville et al. (2004). 
The right panel of their figure 3 shows that the root variance 
of dark matter at $z=6$ for our survey volume 
($1.1 \times 10^6$ Mpc$^3$) is $0.033$ 
in the $\Lambda=0.7$ Cold Dark Matter model.
Assuming a one-to-one correspondence between LBGs and dark haloes, 
we find from the left panel of the same figure that the bias 
parameter of dark haloes at $z=6$ with the same number density 
as of our LBGs ($1.5 \times 10^{-5}$ Mpc$^{-3}$) is 7.3. 
The cosmic variance is thus predicted to be about 
$25\% (=0.033 \times 7.3)$.

\begin{figure}
  \hspace{-45pt}
  \FigureFile(120mm,120mm){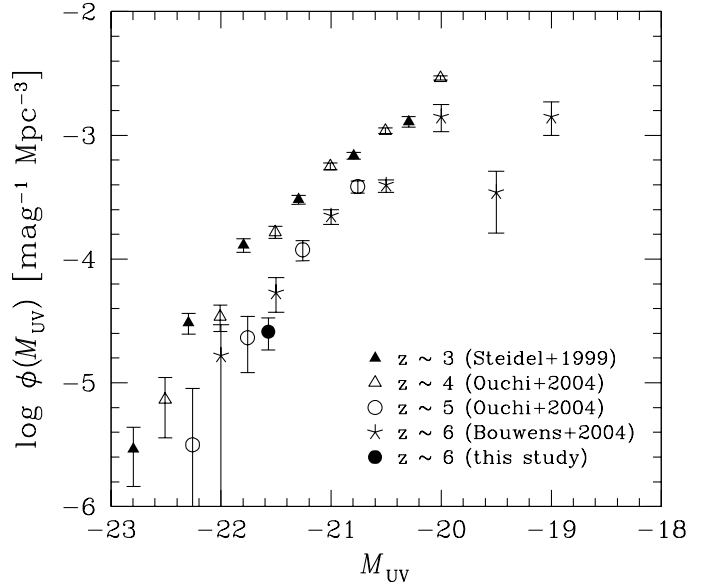}
  \vspace{-30pt}
  \caption{
         FUV luminosity functions of LBGs 
         at $z \sim 3$ (filled triangles: Steidel et al. 1999), 
         $z \sim 4$ (open triangles: Ouchi et al. 2004a), 
         $z \sim 5$ (open circles: Ouchi et al. 2004a), 
         and $z \sim 6$ (stars: Bouwens et al. 2004a; 
         filled circle: this study).
    \label{fig:lf}}
\end{figure}

We plot in Figure \ref{fig:lf} the FUV luminosity function 
measurements for LBGs at $z \sim 3$ (Steidel et al. 1999), 
$z \sim 4$ and $5$ (Ouchi et al. 2004a from shallower SDF data), 
and $z \sim 6$ (Bouwens et al. 2004a and this study). 
The number density of faint $z \sim 6$ LBGs detected in the HUDF 
(Yan \& Windhorst 2004; Bunker et al. 2004) appears 
to be consistent with Bouwens et al.'s (2004a) luminosity function.
We focus on bright ($M \lesssim -21$) LBGs 
for which our measurement is available.
An interesting trend is seen that bright LBGs are less numerous 
at higher redshift, although the difference between 
$z\sim 5$ and 6 may not be so significant.
The number density of bright LBGs drops by a factor of 
about five from $z\sim 3$ to $\sim 6$.

The observed drop of the number density could be due to
either dust absorption or a real decrease in LBGs.
Adelberger \& Steidel (2000) found no significant correlation 
between apparent magnitude and $E(B-V)$ for LBGs at $z \sim 3$. 
A similar result was obtained by Ouchi et al. (2004a) 
for $z \sim 4$ LBGs. 
These results imply that apparently brighter LBGs will 
on average have brighter intrinsic luminosities, 
although there is a large scatter between intrinsic and apparent 
luminosities.
Ouchi et al. (2004a) also found that the $E(B-V)$ distribution 
does not change over $z \sim 3$ and $z \sim 4$.
If the dust properties found in $z \sim 3$ -- 4 LBGs apply to 
$z \ge 5$ LBGs as well, then we can interpret Figure \ref{fig:lf} 
as suggesting that the number density of intrinsically 
bright LBGs decreases with redshift.
Another possibility is that the observed decline 
in bright LBGs is due to an increase in the dust absorption of 
bright LBGs with redshift; 
in this case, the number density of intrinsically bright LBGs 
may not necessarily drop with redshift.
However, there is so far no observation showing evolution 
in dust properties of LBGs with redshift.
In \S 5.3, we discuss the evolution of the FUV luminosity 
density of bright galaxies over $0 \lesssim z \lesssim 6$, 
assuming that the dust properties are unchanged over this 
redshift range.

%
%

\subsection{Star Formation Rates}

We calculate the star formation rates of the 12 LBG candidates 
from their FUV continuum luminosity 
using the formula given in Madau et al. (1998), 
which assumes a Salpeter IMF with $0.1 M_\odot < M < 125 M_\odot$.
In the calculation, all the candidates are assumed to 
be located at $z=5.9$. 
The results are given in Table \ref{tab:lbgs}.
The star formation rates of the 12 candidates 
are found to be $\simeq 20$ -- 30 $M_\odot$ yr$^{-1}$, 
with an average of 23 $M_\odot$ yr$^{-1}$.
The amount of dust extinction for the candidates cannot be 
estimated from our data, but if it is similar to 
the typical value for $z \sim 3$ -- $4$ LBGs, $E(B-V) \simeq 0.15$, 
then the extinction-corrected star formation rate of 
each candidate will be close to $1 \times 10^2 M_\odot$ yr$^{-1}$.
If these candidates maintain the observed star formation rates 
for about 1 Gyr since $z=6$, they will have stellar masses of 
$\gtsim 3 \times 10^{10} M_\odot$ at $z=3$. 
This value is comparable to the observed stellar masses of 
bright $z \sim 3$ LBGs (Papovich et al. 2001; Shapley et al. 2001).

%
%

\subsection{Evolution of the FUV Luminosity Density}

We measure the completeness-corrected FUV luminosity density, 
$\rho_{\rm L}$, 
contributed from LBGs brighter than $M_{\rm FUV} = -21.3$ as:
\begin{equation}
\rho_{\rm L}(<-21.3) 
  \ [{\rm ergs}\ {\rm s}^{-1} {\rm Hz}^{-1} {\rm Mpc}^{-3}]
= \sum_{i=1}^{12} {L(m_i)\over{V_{\rm eff}(m_i)}}
\end{equation}
where $L(m_i)$ is the FUV luminosity (ergs s$^{-1}$ Hz$^{-1}$)
corresponding to apparent magnitude $m_i$.
The absolute magnitude $M_{\rm FUV} = -21.3$ is close to 
$M_{\rm FUV}^\star$ of $z \sim 3 - 4$ LBGs.
We find $\rho_{\rm L}(<-21.3) = (2.8 \pm 0.8) \times 10^{24}$ 
ergs s$^{-1}$ Hz$^{-1}$ Mpc$^{-3}$, 
which is equivalent to a star formation rate density 
of $(3.5 \pm 1.0) \times 10^{-4} M_\odot$ yr$^{-1}$ Mpc$^{-3}$.
If we integrate the luminosity function of Bouwens et al. (2004a) 
over $M_{\rm FUV} < -21.3$, we obtain 
$6.3 \times 10^{-4} M_\odot$ yr$^{-1}$ Mpc$^{-3}$.
Stanway et al. (2003, 2004) estimated the star formation rate 
of $z \sim 6$ LBGs brighter than $z=25.6$ 
in each of the two GOODS fields.
They found 
$(6.7 \pm 2.7) \times 10^{-4} M_\odot$ yr$^{-1}$ Mpc$^{-3}$ 
for the GOODS-South and 
$(5.4 \pm 2.2) \times 10^{-4} M_\odot$ yr$^{-1}$ Mpc$^{-3}$ 
for the GOODS-North.
These values are higher than ours, 
mainly reflecting the higher number density of bright LBGs selected 
in the GOODS fields than in the SDF as discussed in \S 5.1.

As will be discussed below, our sample shows that the FUV luminosity 
density of bright ($M_{\rm FUV} < -21.3$) LBGs at $z\sim 6$ 
is 11 times lower than that at $z \sim 3$.
This ratio is reduced to 6 -- 8 if Stanway et al.'s (2003, 2004) 
estimates are adopted. 
In the same GOODS fields, Giavalisco et al. (2004) found a factor 
of 3.5 decrease from $z \sim 3$ to $z \sim 6$ for $z_{850} < 26$ LBGs.
The smaller decreasing factor found by Giavalisco et al. (2004) is 
probably due to the fainter limiting magnitude of their sample, 
since the evolution of FUV luminosity density tends to be milder 
for fainter galaxies (see below).

\begin{figure}
  \hspace{-45pt}
  \FigureFile(120mm,120mm){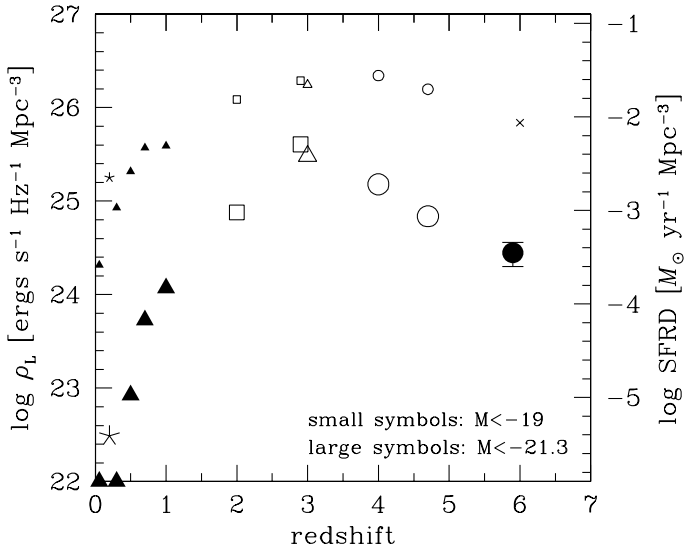}
  \vspace{-30pt}
  \caption{
         Evolution of the FUV luminosity density.
         The large and small symbols indicate the luminosity 
         density contributed from galaxies brighter than 
         $M_{\rm FUV}=-21.3$ and $M_{\rm FUV}=-19.0$, respectively.
         The filled circle with error bars show our measurement, 
         and the other symbols indicate data taken from the literature:
         cross: Bouwens et al. (2004a), 
         open circles: Ouchi et al. (2004a), 
         open triangles: Steidel et al. (1999), 
         open squares: Arnouts et al. (2005; HDF), 
         filled triangles : Arnouts et al. (2005; GALEX), 
         stars: Sullivan et al. (2000).
         The two filled triangles at the bottom 
         imply that the actual measurements are lower than 
         $1 \times 10^{22}$ ergs s$^{-1}$ Hz$^{-1}$ Mpc$^{-3}$.
    \label{fig:lden}}
\end{figure}

We examine how the luminosity densities 
for bright galaxies and all galaxies evolve over $0 \le z \le 6$, 
and whether or not there is a remarkable difference 
in their evolutionary behavior.
Figure \ref{fig:lden} shows the observed evolution of FUV 
luminosity density over $ 0 \le z \le 6$. 
The large symbols indicate the luminosity density 
contributed from bright ($M_{\rm FUV}<-21.3$) galaxies, 
$\rho_{\rm L}(<-21.3)$, while the small symbols are for 
the luminosity density from galaxies down to a faint limit 
of $M_{\rm FUV} = -19.0$, $\rho_{\rm L}(<-19.0)$. 
We use $\rho_{\rm L}(<-19.0)$ as a substitute for 
the total FUV luminosity density in the universe, 
assuming that $\rho_{\rm L}(<-19.0)$ traces the relative change 
in the total luminosity density acceptably, 
although the contribution from galaxies fainter than $-19$ mag 
is uncertain and may be considerably large.
The filled circle indicates our measurement, 
while the other kinds of symbols represent previous measurements 
taken from the literature.
We use Bouwens et al.'s (2004a) luminosity function 
to calculate $\rho_{\rm L}(<-19.0)$ at $z=6$.
The vertical axis on the right-hand side of Figure \ref{fig:lden} 
shows corresponding star formation rate density.
Dust extinction correction has not been applied to any measurements.
The star formation rate density calculated from 
$\rho_{\rm L}(<-19.0)$ multiplied with factor 10 
is roughly consistent with recent measurements 
of the extinction-corrected total star formation rate density 
(Hopkins 2004).

We find that $\rho_{\rm L}(<-21.3)$ evolves drastically 
with redshift. 
It increases by an order of magnitude from $z\sim 6$ to $\sim 3$, 
has a rather sharp peak at $z \sim 3$, and then drops 
by (more than) three orders of magnitude to the present epoch.
The evolution of $\rho_{\rm L}(<-19.0)$, on the other hand, 
is much milder. 
Over $1 \le z \le 6$, $\rho_{\rm L}(<-19.0)$ is roughly 
constant within a factor of about five, and then 
it decreases by ten times from $z\sim 1$ to the present epoch.
As a result,
$\rho_{\rm L}(<-21.3)/\rho_{\rm L}(<-19.0)$ reaches a maximum 
around $z=3$.

The remarkably different evolution of 
$\rho_{\rm L}(<-21.3)$ and $\rho_{\rm L}(<-19.0)$ probably 
reflects a change in $M_{\rm FUV}^\star$ with time.
It has been found that $M_{\rm FUV}^\star$ fades by 3 magnitude 
from $z=3$ ($\simeq -21$) to $z=0$ ($\simeq -18$) 
(Arnouts et al. 2005). 
A faint $M_{\rm FUV}^\star$ is suggested for $z \sim 6$ LBGs 
since no candidate brighter than $z_{\rm R}=24.8$ is detected.
Since $M_{\rm FUV}=-21.3$ is comparable to or brighter than 
$M_{\rm FUV}^\star$ at any redshift, the luminosity function 
has always a sharp, exponential shape around $M_{\rm FUV}=-21.3$. 
Thus, $\rho_{\rm L}(<-21.3)$ is much more sensitive 
to the change in $M_{\rm FUV}^\star$ than $\rho_{\rm L}(<-19.0)$.

An absolute magnitude of $M_{\rm FUV} = -21.3$ corresponds 
to a relatively high star formation rate of $18 M_\odot$ yr$^{-1}$  
(or a dust-corrected value of $80 M_\odot$ yr$^{-1}$ 
if $E(B-V)=0.15$ is assumed).
Figure \ref{fig:lden} shows that such actively star forming 
galaxies become abundant in a short cosmic time from $z=6$ to 3 
(about 1.2 Gyr) and then disappeared by the present time.
Their contribution to the cosmic star formation 
was the highest at $z \sim 3$.

The drastic evolution of actively star forming galaxies 
(i.e., bright LBGs) found here may be physically interpreted 
as a combination of two competing processes 
in a Cold Dark Matter universe:
the build-up of massive dark haloes with time 
and the decrease in the star formation rate 
per unit dark matter mass (i.e., specific star formation rate) 
in dark haloes with time. 
Hernquist \& Springel (2003) and Springel \& Hernquist (2003) 
have made essentially the same discussion 
to explain the observed cosmic star formation history.

Clustering analyses of $z \sim 3-5$ LBGs have shown 
that dark haloes hosting bright $z \sim 3$ -- $5$ LBGs have 
a total mass of $M_{\rm tot} \sim 10^{12} M_\odot$ 
(e.g., Giavalisco \& Dickinson 2001; 
Moustakas \& Somerville 2002; Ouchi et al. 2004b). 
Let us assume that bright $z \sim 6$ 
LBGs also reside in $\sim 10^{12} M_\odot$ dark haloes.
According to the standard Cold Dark Matter model 
with $\Omega_0=0.3$, $\lambda_0=0.7$, $h=0.7$, and $\sigma_8=0.9$, 
the fraction of matter in the universe confined in dark haloes 
with $M_{\rm tot} \ge 10^{12} M_\odot$ 
increases by an order of magnitude from $z=6$ to 3. 
This is similar to the observed evolution rate in the luminosity  
density of bright LBGs.
Note that the model predicts the fraction of matter 
in $M_{\rm tot} \ge 10^{12} M_\odot$ haloes to increase 
by an order of magnitude from $z=3$ to 0; 
the increase per unit cosmic time is thus more gradual at $z<3$ 
than at $3<z<6$.

The decrease with time in the specific star formation rate 
may also be qualitatively in accord with Cold Dark Matter models.
Dark haloes formed later have a lower mass density, 
since the average mass density within dark haloes formed 
at a given cosmic time is proportional 
to the average mass density of the universe at that time.
It is reasonable to expect that specific star formation rate 
correlates positively with the mean mass density within dark haloes.
Hernquist \& Springel (2003) and Springel \& Hernquist (2003)
have shown that gas cooling becomes insufficient in dark haloes 
formed at lower redshifts because of the decline of the mass density, 
and thus the specific star formation rate 
in dark haloes declines with time.
They found that in their numerical simulations the cosmic 
star formation rate has a peak at $z \sim 5-6$.
There may be, of course, other mechanisms which suppress 
star formation in haloes formed recently, 
such as the truncation of star formation in galaxies 
infalling into galaxy groups and clusters which become 
abundant with time, 
and the decrease in galaxy merging involving star formation.

If we adopt this interpretation, 
the observed evolution of actively star forming galaxies 
implies that the specific star formation rate 
of massive ($M_{\rm tot} \sim 10^{12} M_\odot$) dark haloes 
is roughly unchanged over $3 \lesssim z \lesssim 6$, 
while it drops drastically with time at $z \lesssim 3$ 
for some reason, 
resulting in the star formation activity of the whole massive haloes 
being the highest around $z=3$.
The mild evolution of the total luminosity density 
over $z \sim 3$ and 6 looks reasonable, 
since the dark-matter mass summed over all haloes 
increases by only a factor of two or so from $z=6$ to 3.
On the other hand, the mild evolution at $z \lesssim 3$ may 
be due to a gradual decrease, for some reason, 
in the specific star formation rate in less massive haloes.

Finally, we point out an interesting resemblance of the evolution 
of bright LBGs (i.e., FUV-selected actively star-forming galaxies 
brighter than $-21.3$ mag) 
to that of luminous dusty star-forming galaxies (submm galaxies) 
and QSOs.
Chapman et al. (2003, 2004, 2005) found that the space density of 
luminous dusty star-forming galaxies increased by about an order 
of magnitude from $z \sim 4$ to $z \sim 2$ and then decreased by 
$\sim 10^3$ from $z \sim 2$ to the present time, 
having a sharp peak at $z \simeq 2.2$.
Chapman et al. (2003) have pointed out that luminous dusty 
star-forming galaxies coexist with the peak activity of QSOs.
Recent wide-field QSO surveys have measured accurately 
the luminosity function of bright QSOs at $z \lesssim 6$, finding 
that the space density of bright ($M_{\rm FUV} < -26.7$) QSOs 
increases by more than an order of magnitude from $z \sim 6$ 
to $\sim 3$, shows a maximum around $z=2.5$, and then drops 
by about $10^3$ by the present epoch 
(Croom et al. 2004; Fan et al. 2001, 2004).

These three populations appear to be related with active star 
formation or massive galaxy formation.
Dusty star-forming galaxies, detected in submillimeter 
or radio observations, have star formation rates of 
as high as $\sim 10^{3} M_\odot$ yr$^{-1}$ 
(e.g., Chapman et al. 2003, 2004). 
It is widely believed that QSOs are the result of gas accretion 
onto a supermassive black hole at the center of a galaxy, 
and hence are likely be linked to active star formation in the galaxy  
and, possibly, to bulge formation (Franceschini et al. 1999).
Nearby luminous QSOs are hosted by massive bulge-dominated 
galaxies (e.g., Floyd et al. 2004). 
It has also been found that the host galaxies of luminous QSOs 
at $1<z<2$ are consistent with massive galaxies 
following passive evolution (e.g., Falomo et al. 2004).

At $z \sim 2$ -- $3$, the number density of bright LBGs, 
$\sim 10^{-4}$ Mpc$^{-3}$, is higher than those of 
dusty star-forming galaxies ($\sim 10^{-5}$ Mpc$^{-3}$) 
and QSOs ($\sim 10^{-7}$ Mpc$^{-3}$).
This may imply either that the lifetime of bright LBGs, 
i.e., the duration of star formation of 
$\gtsim 20-30 M_\odot$ yr$^{-1}$, is longer than 
the lifetimes of dusty star-forming phase and QSO phase, 
or that bright LBGs reside in less massive 
(thus more numerous) dark haloes than the other two populations, 
or both.
A weak clustering of $z \sim 3$ LBGs compared with the clustering 
of QSOs (Croom et al. 2005) and dusty star-forming galaxies 
(Blain et al. 2004) at similar redshifts 
could imply that LBGs reside in the least massive 
(but still as massive as $\sim 10^{12}M_\odot$) dark haloes.
The relatively milder evolution in number density at $z>2$ 
for LBGs than for dusty star-forming galaxies may also 
suggest smaller dark halo masses of LBGs.
It is also interesting to note that the peak redshift for bright 
LBGs ($z \sim 3$) is slightly higher than those for dusty 
star-forming galaxies and QSOs ($z \sim 2$ -- $2.5$); 
this might reflect hierarchical evolution of haloes that 
less massive haloes are formed earlier.

The physical relationship between the three populations is not 
yet clear, but the resemblance of their evolution 
suggests that $z \sim 3$ is a remarkable era 
in the cosmic history when massive galaxies  
were being intensively formed.
The cosmic time of $z \sim 3$ also falls in the suggested formation 
epoch of massive early-type galaxies seen at $0 \le z \lesssim 2$ 
(e.g., Bower, Lucey, \& Ellis 1992; 
Kodama et al. 1998; 
Stanford et al. 1998; 
van Dokkum \& Stanford 2003; 
Cimatti et al. 2004; 
McCarthy et al. 2004).

%
%

\section{Conclusions}

We showed that LBGs at $z \sim 6$ are isolated well from 
red galaxies at $z \sim 1$ -- $2$ and Galactic cool stars 
in the $i-z_{\rm R}$ vs $z_{\rm B}-z_{\rm R}$ plane. 
We applied this two-color selection to wide-field 
(767 arcmin$^2$) multicolor data of the Subaru Deep Field, 
selecting 12 candidates for $z\simeq 5.9 \pm 0.3$ LBGs 
down to $z_{\rm R}=25.4$.
This is the most reliable sample of bright $z \sim 6$ LBGs, 
which is robust against both contamination by foreground objects 
and cosmic variance.

The average star formation rate of the 12 candidates is 
23 $M_\odot$ yr$^{-1}$ if dust extinction is not corrected. 
We calculated the normalization of the rest-frame 
far-ultraviolet (FUV: $\simeq 1400$\AA) luminosity function 
of $z\sim 6$ LBGs at $M_{\rm FUV}=-21.6$ to be 
$\phi(-21.6) = (2.6 \pm 0.7) \times 10^{-5}$ mag$^{-1}$ Mpc$^{-3}$. 
The FUV luminosity density contributed from 
LBGs brighter than $M_{\rm FUV} = -21.3$ is 
$(2.8 \pm 0.8) \times 10^{24}$ ergs s$^{-1}$ Hz$^{-1}$ Mpc$^{-3}$ 
after correction for detection completeness, 
which corresponds to a star formation rate density 
of $(3.5 \pm 1.0) \times 10^{-4} M_\odot$ yr$^{-1}$ Mpc$^{-3}$.

We examined the evolution of the FUV luminosity density, 
or equivalently the star formation rate density, 
over $0 \le z \le 6$ by combining our measurement with 
those taken from the literature. 
We found that the FUV luminosity density for bright FUV galaxies 
with $M_{\rm FUV} < -21.3$, i.e., actively star-forming galaxies, 
increases by an order of magnitude 
from $z \sim 6$ to $\sim 3$ and takes a rather sharp peak 
at $z \sim 3$, followed by $10^3$ times decrease toward $z=0$.
On the other hand, the evolution of the {\lq}total{\rq} 
luminosity density was found to be much milder.
As a result, the contribution to the cosmic star formation 
from actively star-forming galaxies is the highest at $z \sim 3$.
The drastic evolution of actively star-forming galaxies 
can be explained by a combination of the build-up of massive 
dark haloes with time and the decrease in the specific star 
formation rate in dark haloes with time.
We also pointed out a resemblance between the evolution of 
bright LBGs, luminous dusty star-forming galaxies, and bright QSOs.
The redshift of $z \sim 3$ seems to be a remarkable era 
in the cosmic history in terms of the formation of massive galaxies.

%
%

\vspace{20pt}
We thank the anonymous referee for useful comments which have 
significantly improved the paper.
We are grateful to the Subaru Telescope staff
for their help in the observations.
M. Ouchi has been supported by the Hubble Fellowship program
through grant HF-01176.01-A awarded by the Space Telescope Science
Institute, which is operated by the Association of Universities
for Research in Astronomy, Inc. under NASA contract NAS 5-26555.

%
%

\clearpage

%
%

\begin{table}
  \caption{Photometric properties of LBG candidates.
    \label{tab:lbgs}
  }
  \begin{center}
    \begin{tabular}{rrrrrrrcrccccc}
\hline
\hline
ID & \multicolumn{3}{c}{$\alpha$ (J2000.0)} &
     \multicolumn{3}{c}{$\delta$ (J2000.0)} &
     $z_{\rm R}$& $z_{\rm B}-z_{\rm R}$ & $i-z_{\rm R}$ & 
     NB816 & $z$  & NB921 & SFR$^{(a)}$ \\ 
\hline
 1&   13 & 23 & 46.55    &   +27 & 27 &  3.9   &
    24.84&$ 0.12$ & 2.41 & 27.04 & 25.96 & 25.88 & 29.2 \\
 2&   13 & 25 & 27.81    &   +27 & 28 & 58.9   &
    24.92&$ 0.65$ & 2.79 & 27.04 & 25.42 & 25.37 & 27.1 \\
 3&   13 & 24 & 39.15    &   +27 & 32 & 13.0   &
    24.94&$ 0.47$ & 2.50 & 27.04 & 26.06 & 26.42 & 26.6 \\
 4&   13 & 23 & 49.72    &   +27 & 45 & 27.4   &
    25.02&$-0.14$ & 1.47 & 26.10 & 26.57 & 26.07 & 24.7 \\
 5&   13 & 24 & 16.91    &   +27 & 13 & 48.3   &
    25.04&$ 0.06$ & 2.28 & 27.04 & 26.01 & 26.16 & 24.2 \\
 6&   13 & 25 & 23.15    &   +27 & 42 & 51.1   &
    25.14&$ 0.26$ & 2.66 & 27.04 & 26.24 & 25.76 & 22.1 \\
 7&   13 & 24 &  9.43    &   +27 & 18 & 30.4   &
    25.15&$ 0.36$ & 1.85 & 25.75 & 25.55 & 25.42 & 21.9 \\
 8&   13 & 23 & 39.77    &   +27 & 31 & 37.4   &
    25.16&$-0.01$ & 1.89 & 27.04 & 25.49 & 25.68 & 21.7 \\
 9&   13 & 24 & 27.13    &   +27 & 16 & 40.1   &
    25.23&$ 0.48$ & 2.22 & 27.04 & 26.13 & 25.95 & 20.4 \\
10&   13 & 23 & 55.87    &   +27 & 32 &  9.3   &
    25.26&$ 0.14$ & 1.61 & 26.05 & 25.73 & 25.94 & 19.8 \\
11&   13 & 24 & 40.01    &   +27 & 43 & 32.3   &
    25.36&$ 0.32$ & 1.93 & 26.95 & 26.19 & 25.87 & 18.1 \\
12&   13 & 25 &  7.62    &   +27 & 45 &  7.4   &
    25.36&$ 0.43$ & 2.49 & 27.04 & 25.99 & 25.83 & 18.1 \\
\hline
    \end{tabular}
  \end{center}
\end{table}

\end{document}